\documentclass[aps,prd,twocolumn,groupedaddress,amssymb,eqsecnum,showpacs,epsfig,nofootinbib]{revtex4}
\pdfoutput=1
\usepackage{graphicx}
\usepackage{bm}
\usepackage{dcolumn}
\usepackage{amsmath}
\usepackage{amsmath}
\usepackage{amssymb}

\renewcommand \thesection{\arabic{section}}

 \newcommand{\nn}{\nonumber}
\numberwithin{equation}{section}
\usepackage{epsfig}
\begin{document}

\title{Direction Dependence of the Deceleration Parameter}

\author{Rong-Gen Cai}
\email{cairg@itp.ac.cn}

\author{Zhong-Liang Tuo}
\email{tuozhl@itp.ac.cn}

\address{Key Laboratory of Frontiers in Theoretical Physics,
    Institute of Theoretical Physics, Chinese Academy of Sciences,
    P.O. Box 2735, Beijing 100190, China}

\date{August, 2011}

%************************************   Abstract   *******************************************%

\vspace*{0.3cm}

\begin{abstract}
\begin{center}
{\bf Abstract}
\end{center}
In this paper we study the possibly existing anisotropy in the
accelerating expansion Universe by use of the full sample of Union2
data. Using the hemisphere comparison method to search for a
preferred direction, we take the deceleration parameter $q_0$ as the
diagnostic to quantify the anisotropy level in the $wCDM$ model. We
find that the maximum accelerating expansion direction is
$(l,b)=(314_{\ +20^{\circ}} ^{\circ-13^{\circ}},28_{\
+11^{\circ}}^{\circ-33^{\circ}})$, with the maximum anisotropy level
of $\Delta q_{0,max}/\bar{q}_0=0.79_{+0.27}^{-0.28}$, and that the
anisotropy is more prominent when only low redshift data
($z\leq0.2$) are used. We also discuss this issue in the $CPL$
parameterized model, showing a similar result.

\end{abstract}

%*********************************************************************************************%

\pacs{98.80.Es, 98.80.Jk}
%98.80.Es -> Observational cosmology
%98.80.Jk -> Mathematical and relativistic aspects of cosmology
\maketitle

%***********************************   Introduction   ****************************************%

\section{\normalsize{Introduction}}
\renewcommand{\thesection}{\arabic{section}}

Since the discovery of cosmic acceleration in the late of last
century~\cite{SNIa}, the type Ia supernovae (SNIa) have become an
important tool in determining the cosmological parameters. Given
certain cosmological models, one usually implements the joint
analysis with SNIa in combination with other observations, such as
large scale structure~\cite{BAO}, the cosmic microwave background
(CMB) radiation~\cite{CMB}, and so on, to give constraints of
cosmological parameters.

Two pillars of modern cosmology are general relativity and the
cosmological principle. It is assumed that Einstein's general
relativity still holds on the cosmic scale, which means that the
evolution of the universe is governed by general relativity. The
cosmological principle~\cite{Weinberg} says that our universe is
homogeneous and isotropic on the cosmic scale. Indeed, the
assumption of homogeneity and isotropy is consistent with currently
accurate data coming from the cosmic microwave background (CMB)
radiation, especially from the Wilkinson Microwave Anisotropy Probe
(WMAP)~\cite{WMAP7}, the statistics of galaxies~\cite{galaxy}, and
the halo power spectrum~\cite{halo}, etc. And current astronomical
observations are in agreement with $\Lambda CDM$ model~\cite{LCDM}.

However, the standard model is also challenged by some
observations~\cite{challenge} (for more details see~\cite{lcdm} and
references therein). Thus it is necessary to revisit the assumption
of the homogeneity and isotropy. As more and more supernovae data
are released~\cite{Union2, Union2.1}, this study becomes possible.

From the theoretical point of view, the anisotropy may arise in some
cosmological models. For example, a vector field may lead to
anisotropy of the universe and gives rise to an anisotropic equation
of state of dark energy~\cite{vector}. Peculiar velocities are also
associated with dipole-like anisotropies, triggered by the fact that
they introduce a preferred spatial direction, and as a result of the
drift motion, observers may find that the acceleration is maximized
in one direction and minimized in the opposite~\cite{peculiar}. By
use of 288 SNIa~\cite{Kessler:2009ys},  Davis {\it et al.} studied
the effects of peculiar velocities on cosmological parameters,
including our own peculiar motion, supernova's motion and coherent
bulk motion, and found that neglecting coherent velocities in the
current sample would cause a systematic shift in the equation of
state of dark energy, with deviation $\Delta w=0.02$~\cite{davis}.
In this paper, we focus on the issue related to the possible
existence of anisotropy, and search for a preferred direction by
using the SNIa Union2 (consisting of 557 SNIa)
data\footnote{http://supernova.lbl.gov/Union/.}. Some previous works
 payed attention to this issue using the SNIa data and found no
statistically significant evidence for anisotropies~\cite{previous}.
Using the Union2, some authors derived the angular covariance
function of the standard candle magnitude fluctuations, searching
for angular scales where the covariance function deviates from 0 in
a statistically significant manner, and no such angular scale was
found~\cite{noevidence}. Yet on the other hand, using the SNIa data
in the frame work of an anisotropy Bianchi type I cosmological model
and in the presence of a dark energy fluid with anisotropy equation
of state, Ref.~\cite{previous1} found that a large level of
anisotropy was allowed both in the geometry of the universe and in
the equation of state of dark energy. Ref.~\cite{previous2}
constructed a ``residual'' statistic, which is sensitive to
systematic shifts in the SNIa brightness, and used this to search in
different slices of redshift for a preferred direction on the sky,
and found that at low redshift ($z<0.5$) an isotropic model was
barely consistent with the SNIa data at $2$-$3\sigma$. In addition,
Ref.~\cite{schwarz} took use of the hemisphere comparison method to
fit the $\Lambda CDM$ model to the supernovae data on several pairs
of opposite hemispheres, and a statistically significant preferred
axis was found. More recently, Antoniou and
Perivolaropoulos~\cite{challenge1} have applied the hemisphere
comparison method to the standard $\Lambda CDM$ model and found that
the hemisphere of maximum accelerating expansion is in the direction
$(l,b)=(309^{\circ -3^{\circ}}_{\
+23^{\circ}},18^{\circ-10^{\circ}}_{\ +11^{\circ}})$ with Union2
data. This result is consistent  with other observations, such as
CMB dipole~\cite{cmbdipole}, CMB quadrupole~\cite{cmbquadrupole},
CMB octopole~\cite{cmboctopole}, large scale velocity
flows~\cite{velocity} and large scale alignment in the QSO optical
polarization data~\cite{quasar}. In these observations, all the
preferred directions appear to be towards the North Galactic
Hemisphere.  They obtained the average direction of the preferred
axes as $(l,b)=(278^{\circ}\pm 26^{\circ},45^{\circ}\pm
27^{\circ})$.

In this paper we study the dependence of this result on the dark
energy model. We consider two models. One is the $wCDM$ model; the
other is a dynamical dark energy model with the $CPL$
parametrization~\cite{CPL}. We take the present value of
deceleration parameter $q_0$ as the diagnostic to quantify the
anisotropy level of two opposite hemispheres.

The paper is organized as follows. In the next section we give a
general introduction to the hemisphere comparison method using $q_0$
to quantify the anisotropy level and we apply it to $wCDM$ model
fitted by the Union2 dataset. In Sec. 3, we give the numerical
results on the preferred directions from the SNIa data with
different slices of redshift. We also give the result for the $CPL$
model. Sec. 4 gives  our conclusions.

%*********************************************************************************************%

\section{\normalsize{Hemisphere Comparison Method Using The Union2 Dataset}}
\renewcommand{\thesection}{\arabic{section}}

Since its first release, the SNIa data have been becoming an
important tool for us to understand the evolution of the universe.
And as time goes on, more accurately determined data will be
released. For example, the future Joint Dark Energy Mission
(JDEM)~\footnote{http://jdem.lbl.gov/.} is aimed to explore the
properties of dark energy and to measure how cosmic expansion has
changed over time. In this paper we take use of the Union2
dataset~\cite{Union2}, which contains 557 type Ia SNIa data and uses
SALT2 for SNIa light-curve fitting, covering the redshift range
$z=[0.015,1.4]$ and including samples from other surveys, such as
CfA3~\cite{cfa}, SDSS-II Supernova Search~\cite{sdss} and high-z
Hubble Space Telescope.

We fit the SNIa data by minimizing the $\chi^2$ value of the
distance modulus. The $\chi_{sn}^{2}$ for SNIa is obtained by
comparing theoretical distance modulus
$\mu_{th}(z)=5\log_{10}[d_L(z)]+\mu_{0}$, where
$\mu_{0}=42.384-5\log_{10}h$ is a nuisance parameter, with observed
$\mu_{ob}$ of supernovae:
\[
\chi_{sn}^{2}=\sum_{i=1}^{557}\frac{[\mu_{th}(z_{i})-\mu_{ob}(z_{i})]^{2}}{\sigma^{2}(z_{i})}.
\]
For a flat FRW cosmological model, one has
\begin{equation}
d_L(z)=(1+z)\int_{0}^{z}\frac{H_0}{H({z}')}d{z}'.
\end{equation}
In the case of $wCDM$ model, we have
\begin{equation}
H^2(z)=H^2_0[\Omega_{m0}(1+z)^3+(1-\Omega_{m0})(1+z)^{3+3w}].
\end{equation}
And the present value of the deceleration parameter can be expressed
as
\begin{equation}
\label{eq2.3}
 q_0=\frac{1}{2}+\frac{3}{2}w(1-\Omega_{m0}).
\end{equation}
In the case of the $CPL$ parametrization, the equation of state of
dark energy is $w=w_0+w_1\frac{z}{1+z}$. Accordingly one obtains
\begin{eqnarray}
H(z)&=&H_0^2[\Omega_{m0}(1+z)^3+(1-\Omega_{m0})(1+z)^{3+3w_0+3w_1}\nn\\
&&\exp{(-3w_1 z/(1+z))}],\\
\label{eq2.5}
q_0&=&\frac{1}{2}+\frac{3}{2}w_0(1-\Omega_{m0}).\end{eqnarray}

Since $\mu_0$ is a nuisance parameter, we can eliminate the effect
of $\mu_{0}$ in the following way. We expand $\chi_{sn}^{2}$ with
respect to $\mu_{0}$ \cite{Nesseris:2005ur}:
\begin{equation}
\chi_{sn}^{2}=A+2B\mu_{0}+C\mu_{0}^{2},\label{eq:expand}\end{equation}
 where \begin{eqnarray*}
A & = & \sum_{i}\frac{[\mu_{th}(z_{i};\mu_{0}=0)-\mu_{ob}(z_{i})]^{2}}{\sigma^{2}(z_{i})},\\
B & = & \sum_{i}\frac{\mu_{th}(z_{i};\mu_{0}=0)-\mu_{ob}(z_{i})}{\sigma^{2}(z_{i})},\\
C & = & \sum_{i}\frac{1}{\sigma^{2}(z_{i})}.\end{eqnarray*}
 Eq.~(\ref{eq:expand}) has a minimum as \[
\widetilde{\chi}_{sn}^{2}=\chi_{sn,min}^{2}=A-B^{2}/C,\] which is
independent of $\mu_{0}$. In fact, it is equivalent to performing a
uniform marginalization over $\mu_{0}$, the difference between
$\widetilde{\chi}_{sn}^{2}$ and the marginalized $\chi_{sn}^{2}$ is
just a constant \cite{Nesseris:2005ur}. We will adopt
$\widetilde{\chi}_{sn}^{2}$ as the goodness of fitting between
theoretical model and SNIa data.

The directions to the SNIa we use here are given in
Ref.~\cite{noevidence}, and are described in the equatorial
coordinates (right ascension and declination). In order to use the
hemisphere comparison method, we need to convert these coordinates
to usual spherical coordinates $(\theta,\phi)$, and finally to the
galactic coordinates $(l,b)$~\cite{convert}.

We apply the hemisphere comparison method to find the possibly
existing preferred axis. This method was first proposed in
Ref.~\cite{schwarz}, and further developed in
Ref.~\cite{challenge1}, while the difference between them is the
choice of the direction which separates the data into two subsets.
By rotating the poles at $l\in[0^{\circ}, 180^{\circ}]$ and
$b\in[-90^{\circ}, 90^{\circ}]$ in every $1^{\circ}$ step, the
original method uses definite number of directions. With the
developed method, one can have more random directions compared with
the original one, by setting the repeating times (see below).
Therefore we will take the developed method. The hemisphere
comparison method involves the following steps.

1. Generate a random direction
\begin{equation}
\hat{n}=(\cos\phi \sin\theta,\sin\phi\sin\theta,\cos\theta )
\end{equation}
where $\phi\in[0,2\pi)$ and $\theta\in[0,2\pi]$ are random numbers
with uniform probability distribution. Note that the uniform
probability distribution of these two variables $(\phi, \theta)$
will guarantee that every direction will be generated with the same
probability, other priors on the distribution of $\phi$ or $\theta$
may lead to a priori preferred direction.

2. Split the dataset under consideration into two subsets according
to the sign of the value $\hat{n}\cdot\hat{n}_{dat}$, where
$\hat{n}_{dat}$ is the unit vector describing the direction of each
SNIa in the dataset. Then we divide the data in two opposite
hemispheres, denoted by up and down, respectively.

3. Find the best-fitting values of $(\Omega_{m0},w)$ ($(\Omega_{m0},
w_0)$ for $CPL$) on each hemisphere. Naturally, one can take use of
the deceleration parameter $q_0$ to quantify the anisotropy level
through the normalized difference
\begin{equation}
\label{eq2.8} \frac{\Delta
q_0}{\bar{q}_0}=2\frac{q_{0,u}-q_{0,d}}{q_{0,u}+q_{0,d}}
\end{equation}
where the subscripts $u$ and $d$ denote the up and down hemispheres,
respectively. Note that the deceleration parameter is a good
diagnostic to quantify the anisotropy level. The hemisphere with
larger $q_0$ is expanding slower than the opposite. Larger
normalized difference in one direction means that the anisotropy
level in this direction is more notable.

4. Repeat 400 times from step 1 to step 3, and find the maximum
normalized difference defined by (\ref{eq2.8}) for the Union2 data,
thus one can obtain the corresponding direction of maximum
anisotropy.

As illustrated in~\cite{challenge1}, in order to maximize the
efficiency, the number of directions should be no less than the
number of data points of SNIa on each hemisphere. The reason for
this is that changing the direction of an axis, does not change the
corresponding $\frac{\Delta q_0}{\bar{q}_0}$ until a data point is
crossed by the corresponding equator line. Such a crossing is
expected to occur when the direction of an axis changes by the
approximately  mean angular separation between data points. Thus,
using more axes than the number of data points in a hemisphere does
not improve the accuracy of the determination of the maximum
anisotropy direction. Given that the number of data points per
hemisphere for the Union2 dataset is about 280, we have used 400
axes in our analysis, well above the value of 280.

%*********************************************************************************************%

\section{\normalsize{Results}}
\renewcommand{\thesection}{\arabic{section}}

Following the steps introduced in Sec.II, one can obtain the
best-fitting values of $q_0$ in each direction, and then find the
maximum anisotropy direction. But in order to obtain the $1\sigma$
errors of the maximal anisotropic direction and the anisotropy
level, we need to get the errors of the parameters
($(\Omega_{m0},w)$ for $wCDM$ and $(\Omega_{m0}, w_0)$ for $CPL$),
which would propagate to $q_0$ and then to $\frac{\Delta
q_0}{\bar{q}_0}$, because $q_0$ is determined by $w$ and
$\Omega_{m0}$ [see (\ref{eq2.3})] for $wCDM$ model or $w_0$ and
$\Omega_{m0}$ [see (\ref{eq2.5})] for CPL model. The analysis here
is performed by using the Monte Carlo Markov Chain in the
multidimensional parameter space to derive $1\sigma$ errors on each
hemispheres in the maximum anisotropic direction. Accordingly, the
$1\sigma$ deviation from the maximum anisotropy level can be
expressed as $\frac{\Delta q_0}{\bar{q}_0}=\frac{\Delta
q_{0,max}}{\bar{q}_{0,max}}\pm \sigma_{\delta q}$, and
correspondingly the direction axes with $1\sigma$ error can also be
obtained.

\begin{center}
 \begin{figure}
 \includegraphics[width=3.5in]{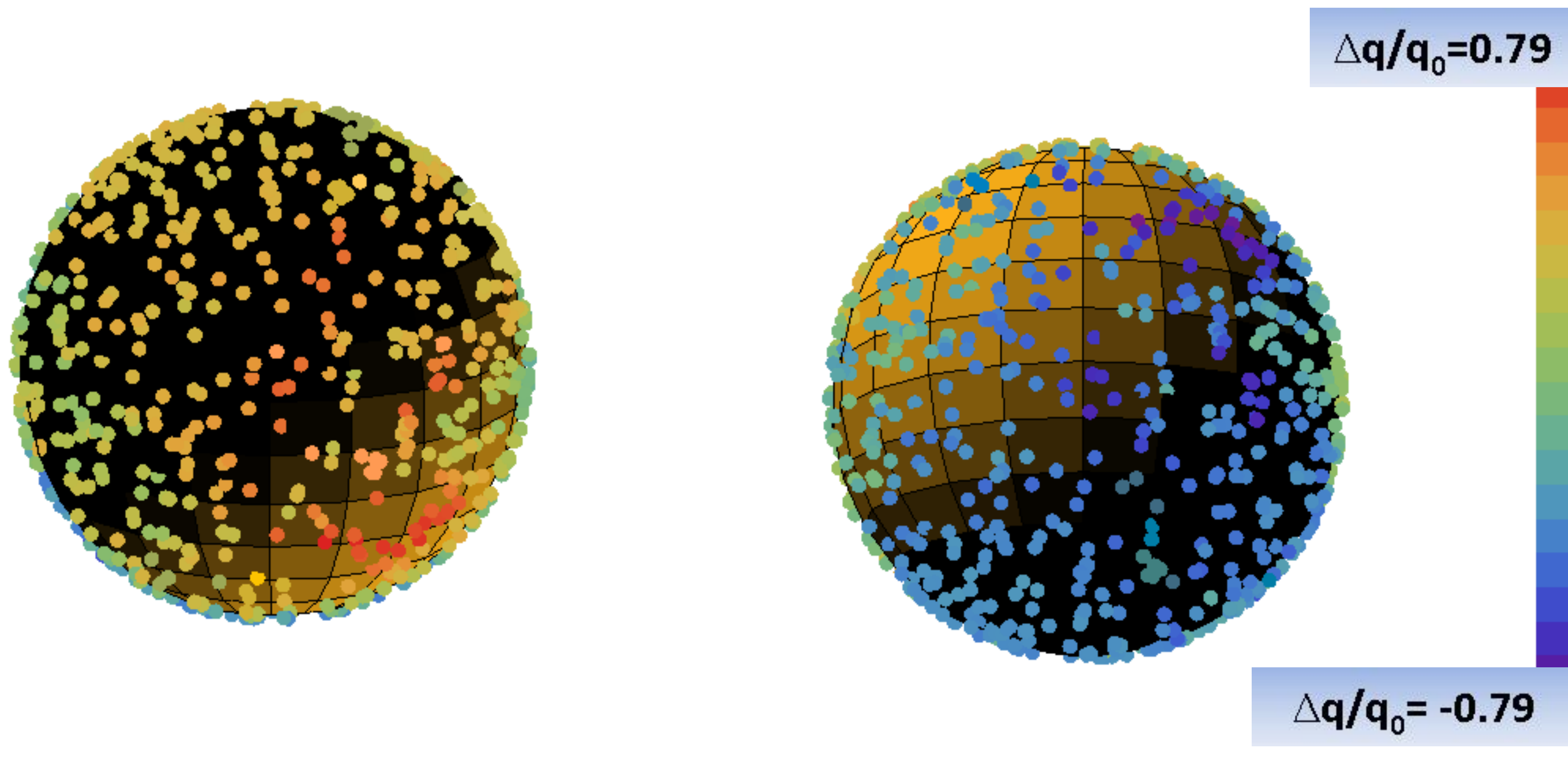}
 \caption{\label{fig:pic}The dots on the unit sphere
colored according to the sign and magnitude of the anisotropy level
are the directions of the random axes. The hemisphere shown on the
left panel is the one corresponding to smaller accelerations with
the preferred axis $(l,b)=(134^{\circ},-28^{\circ})$ [$q_0=-0.40$ in
this direction], while the right corresponds to the one with larger
accelerations and the preferred axis
 $(l,b)=(314^{\circ},28^{\circ})$
and $q_0=-0.92$. }

\end{figure}
 \end{center}

Further we explore the possible redshift dependence of the
anisotropy. We implement a redshift tomography of the data and take
the same procedure as before for all the following redshift slices:
0-0.2, 0-0.4, 0-0.6, 0-0.8, 0-1.0. Our results for $wCDM$ model are
summarized in Table \ref{tab:result}. And we also show the result
for the full sample of Union2 $(0<z\leq1.4)$ in Figure
\ref{fig:pic}.

\begin{table}[!h]
\begin{tabular}{|c|c|c|c|c|}
\hline ${\rm redshift\, range}$  & $l[\rm degree]$  & $b[\rm
degree]$ & $\frac{\Delta q_0}{\bar{q}_0}$ \tabularnewline \hline
\hline $0-0.2$ & $332^{-35}_{+3}$ & $-29^{-11}_{+42}$ &
$3.103^{-0.54}_{+0.50}$\tabularnewline \hline \hline $0-0.4$ &
$319_{+20}^{-12}$ & $-28_{+40}^{-25}$  &
$1.00_{+0.47}^{-0.51}$\tabularnewline \hline \hline $0-0.6$ &
$300_{+33}^{-16}$ & $-16_{+24}^{-10}$ &
$0.94_{+0.37}^{-0.39}$\tabularnewline \hline \hline $0-0.8$ &
$309_{+8}^{-31}$ & $21_{+28}^{-41}$ &
$0.91_{+0.30}^{-0.29}$\tabularnewline \hline \hline $0-1.0$ &
$311_{+15}^{-20}$ & $15_{+17}^{-19}$ &
$0.85_{+0.28}^{-0.28}$\tabularnewline \hline \hline $0-1.4$ &
$314_{+20}^{-13}$ & $28_{+11}^{-33}$ & $0.79_{+0.27}^{-0.28}$
 \tabularnewline \hline
\end{tabular}
\tabcolsep 0pt \caption{\label{tab:result}Directions of maximum
anisotropy for several redshift ranges of the Union2 data fitting
with the $wCDM$ model. The last column corresponds to the $1\sigma$
errors propagated from $\Omega_{m0}$ and $w$, which are obtained by
 MCMC method.} \vspace*{5pt}
\end{table}

The redshift tomography analysis here shows that the preferred axes
are all located in a relatively small part of the North Galactic
Hemisphere [around $(l,b)=(314_{\
+20^{\circ}}^{\circ-13^{\circ}},28_{\
+11^{\circ}}^{\circ-33^{\circ}})$], which is consistent with the
result in~\cite{challenge1} at $1\sigma$ confidence level, thus
indicates that under the assumption of $wCDM$ model, the universe
has a maximum acceleration direction. For different redshift slices,
there are slight differences in the direction of preferred axes, and
the difference between the two opposite hemispheres is extremely
obvious for the low-redshift slice $(z\leq0.2)$.

Here we also give the best-fitting parameters of the $wCDM$ model in
Table \ref{tab:compare}, where the subscripts $u$ and $d$ denote the
up and down hemispheres. For the case with full SNIa data, see the
last row in Table 2. Note that in the $\Lambda CDM$ model,
$\Omega_{m0}^u=0.30$ and $\Omega_{m0}^d=0.19$~\cite{challenge1}.
\begin{table}[!h]
\begin{tabular}{|c|c|c|c|c|}
\hline ${\rm redshift\, range}$  & $\Omega_{m0}^{u}$  &
$\Omega_{m0}^{d}$  & $w^{u}$ & $w^{d}$ \tabularnewline \hline \hline
$0-0.2$ & $0.45$ & $0.10$ & $-0.80$ & $-0.51$\tabularnewline \hline
\hline $0-0.4$ & $0.44$ & $0.11$  & $-1.93$ & $-0.65$\tabularnewline
\hline \hline $0-0.6$ & $0.41$ & $0.11$ & $-1.91$ &
$-0.70$\tabularnewline \hline \hline $0-0.8$ & $0.37$ & $0.12$ &
$-1.68$ & $-0.67$\tabularnewline \hline \hline $0-1.0$ & $0.37$ &
$0.18$ & $-1.68$ & $-0.76$\tabularnewline \hline \hline $0-1.4$ &
$0.35$ & $0.22$ & $-1.46$ & $-0.77$
 \tabularnewline \hline
\end{tabular}
\tabcolsep 0pt \caption{\label{tab:compare}Best-fitting parameters
of the $wCDM$ model for different redshift slices on the opposite
hemispheres in the direction of the maximum anisotropy, where $u$
denotes the hemisphere corresponding to larger accelerations, while
$d$ denotes the opposite hemisphere.} \vspace*{5pt}
\end{table}

Note that $\Omega_{m0}$ is used as the diagnostic to the anisotropy
level in~\cite{challenge1}. Here we use $q_0$ as the diagnostic to
the anisotropy level. We found here that $q_0$ is more sensitive to
the errors of the data, since $q_0$ is relative to the second order
derivative of the luminosity distance. And also because more
parameters involved in the analysis will introduce more
uncertainties when we try to determine the error of $q_0$, therefore
the errors here are larger than those in~\cite{challenge1}. For
example, the dots colored according to the sign and magnitude of the
anisotropy level will be more scattered on the unit sphere, as shown
in Figure \ref{fig:pic}. But the meaning of using $q_0$ is obvious
that the hemisphere with larger $q_0$ is expanding slower than the
opposite, and that the direction with normalized difference
$\frac{\Delta q_0}{\bar{q}_0}$ means that the anisotropy level in
this direction is more notable. What's more, it is clear from
Figure~\ref{fig:pic} that the sphere is divided into two distinct
hemispheres, one with smaller accelerations and the other with
larger accelerations.

With the same procedure  we also study this issue for a dynamical
dark energy model with the $CPL$ parametrization. By use of all
Union2 data points, we find that the direction of preferred axis is
$(l,b)=(309_{\ +30^{\circ}}^{\circ -23^{\circ}},21_{\
+35^{\circ}}^{\circ-26^{\circ}})$, and the corresponding maximum
anisotropy  is $\frac{\Delta
q_{0,max}}{\bar{q}_{0,max}}=0.76_{+0.41}^{-0.46}$. The result shows
that it is not much different from the case of the $wCDM$ model.
This means that the best-fitting values of the preferred direction
is not very sensitive to the dark energy models.

%***********************************   Conclusions   *****************************************%

\section{\normalsize{Conclusions}}
\renewcommand{\thesection}{\arabic{section}}

From some astronomical observations and some theoretical models of
the universe, there seemingly exists some evidence for a
cosmological preferred axis~\cite{lcdm}. If such a cosmological
preferred axis indeed exists, one has to seriously consider an
anisotropic cosmological model as a realistic model, instead of the
FRW universe model.

In this paper we investigated the existence of anisotropy of the
universe by employing the hemisphere comparison method and the
Union2 SNIa dataset and  found this preferred direction. We used the
present value of the deceleration parameter $q_0$ to quantify the
anisotropy level of the two hemispheres. For the $wCDM$ model, the
preferred direction is $(l,b)=(314_{\
+20^{\circ}}^{\circ-13^{\circ}},28_{\
+11^{\circ}}^{\circ-33^{\circ}})$, and  $\frac{\Delta
q_0}{\bar{q_0}}=0.79 ^{-0.28}_{+0.27}$. While in the case of CPL
model, the direction of preferred axis is $(l,b)=(309_{\
+30^{\circ}}^{\circ -23^{\circ}},21_{\
+35^{\circ}}^{\circ-26^{\circ}})$, and correspondingly the maximum
anisotropy level is $\frac{\Delta
q_{0,max}}{\bar{q}_{0,max}}=0.76_{+0.41}^{-0.46}$. Comparing with
the result given in~\cite{challenge1}, where the $\Lambda CDM$ model
is employed, our results are basically in agreement. This means that
the best-fitting preferred direction is not much sensitive to the
dark energy models.

Finally let us notice that it can be seen from Table 1 that the
result is weakly dependent on redshift if the redshift tomography
analysis is employed.

%*********************************   Acknowledgments   ***************************************%

\begin{acknowledgments}
\vspace*{1cm}
 We would like to thank Prof. Z. K. Guo for useful
discussions and valuable comments. This work was supported in part
by the National Natural Science Foundation of China (No. 10821504,
No. 10975168, No.11035008 and No.11075098),  by the Ministry of
Science and Technology of China under Grant No. 2010CB833004 and by
a grant from the Chinese Academy of Sciences.
\end{acknowledgments}

%**********************************   Bibliography   *****************************************%

\end{document}